\documentclass[sigconf]{acmart}
\AtBeginDocument{%
  }

\setcopyright{acmlicensed}
\copyrightyear{2025}
\acmYear{2025}
\acmDOI{XXXXXXX.XXXXXXX}
\acmBooktitle{The ACM International Conference on the Foundations of Software Engineering (FSE'25), June 23--27, 2025, Trondheim, Norway}
\acmISBN{978-1-4503-XXXX-X/2018/06}

\usepackage{xcolor}
\definecolor{APA_stats}{RGB}{100, 100, 120}
\newcommand{\APAstats}[2]{\textcolor{APA_stats}{(M=#1, SD=#2)}}

\begin{document}

\title{A Multi-agent Onboarding Assistant based on Large Language Models, Retrieval Augmented Generation, and Chain-of-Thought}

\author{Andrei-Cristian Ionescu}
\affiliation{%
  \institution{Delft University of Technology}
  \city{Delft}
  \country{The Netherlands}
}
\email{a.c.ionescu-1@student.tudelft.nl}

\author{Sergey Titov}
\affiliation{%
  \institution{JetBrains Research}
  \city{Amsterdam}
  \country{The Netherlands}
}
\email{sergey.titov@jetbrains.com}

\author{Maliheh Izadi}
\orcid{0000-0001-5093-5523}
\affiliation{%
  \institution{Delft University of Technology}
  \city{Delft}
  \country{The Netherlands}
}
\email{m.izadi@tudelft.nl}


\begin{abstract}
Effective onboarding in software engineering is crucial but difficult due to the fast-paced evolution of technologies. Traditional methods, like exploration and workshops, are costly, time-consuming, and quickly outdated in large projects.
We propose the \textit{Onboarding Buddy} system, which leverages large language models, retrieval augmented generation, and an automated chain-of-thought approach to improve onboarding. It integrates dynamic, context-specific support within the development environment, offering natural language explanations, code insights, and project guidance. Our solution is agent-based and provides customized assistance with minimal human intervention. 
Our study results among the eight participants show an average helpfulness rating of \APAstats{3.26}{0.86} and ease of onboarding at \APAstats{3.0}{0.96} out of four.
While similar to tools like GitHub Copilot, Onboarding Buddy uniquely integrates a chain-of-thought reasoning mechanism with retrieval-augmented generation, tailored specifically for dynamic onboarding contexts. 
While our initial evaluation is based on eight participants within one project, we will explore larger teams and multiple real-world codebases in the company to demonstrate broader applicability.
Overall, Onboarding Buddy holds great potential for enhancing developer productivity and satisfaction. 
Our tool, source code~\cite{onboardingbuddy2025software}, and demonstration video~\cite{onboardingbuddy2025video} are publicly available.
\end{abstract}



\maketitle

\section{Introduction}
Software engineering is a dynamic domain; the frequency of learning new technologies, frameworks, and tools for software developers is very high. 
The newcomers also need to adapt to the codebase, project requirements, and team culture rapidly. 
The domain itself is constantly changing with emerging new trends that may require developers to continuously update their skills~\cite{ju2021case,britto2017onboarding}. 
Such challenges may be met through the proper designing of an onboarding system, but such systems are expensive and time-consuming to put into effect for large projects. 

Traditional methods, such as documentation and workshops, may no longer be useful, codebase exposure being a more important factor in understanding a project~\cite{ernst2023documentation}. With recent developments in large language models (LLMs)~\cite{al2023extending,izadi2024language}, we can quickly forward relevant advice to each developer~\cite{MacNeil2023Experiences}. However, managing and implementing this guidance in an effective manner remains a challenge~\cite{sergeyuk2024design}. 

To address this issue, we introduce the \textit{Onboarding Buddy}. 
Onboarding Buddy automates the onboarding process by automatically generating detailed, project-specific explanations to users' questions~\cite{deljouyi2024leveraging}. Doing this cuts down the heavy mentoring needed from other colleagues. The system aims to assist new developers in going faster through the codebase and its logic but also to minimize the frustration and stress associated with onboarding. The system integrates components such as retrieval augmented generation (RAG) \cite{lewis2021retrievalaugmentedgenerationknowledgeintensivenlp}, and automated chain-of-thought (CoT)~\cite{wei2023chainofthoughtpromptingelicitsreasoning} with a large language model to present an end-to-end solution, making it more effective for onboarding new team members.

Unlike existing general-purpose LLM coding tools, such as GitHub Copilot, which are general in their assistance, Onboarding Buddy has been designed to tackle the particular challenges of onboarding into new codebases. Its scratchpad mechanism keeps track of the intermediate steps in reasoning and reduces the noise that arises in multi-agent interactions. This is an explicit design choice to offer more explainability and adaptability during onboarding tasks. 

To evaluate the Onboarding Buddy, eight users performed onboarding tasks related to setting up their development environment and completing tasks on a backend application. 
We chose a closed-source project to avoid data contamination with models' opaque training data.
Seven out of eight participants successfully completed all tasks, with only one participant making minor coding errors on one of the tasks. After completing the tasks, we asked participants to rate the perceived helpfulness and ease of onboarding on a scale from one to four. As a result, over all the participants and tasks we obtained a high mean rating of perceived helpfulness of \APAstats{3.26}{0.86}. 
Moreover, there was a relatively good mean ease-of-onboarding rating of \APAstats{3.0}{0.96}.
Our study has the following contributions:
\begin{itemize}
    \item Introducing a novel LLM-powered onboarding tool that leverages dynamic and contextualized onboarding experiences.
    \item Presenting its agent-centric architecture, including specialized agents like the Onboarding Agent,
    \item Providing the implementation of RAG and automated CoT process, which enhance reasoning and retrieval, and
    \item Conducting an empirical evaluation of the IntelliJ Onboarding Plugin with eight participants. 
\end{itemize}

\section{Related Work}

Onboarding in software companies involves integrating new developers into the team and project by helping them understand the codebase, architecture, and business logic.

According to Ju et al.,~\cite{ju2021case}, effective onboarding is directly related to productivity, job satisfaction, and outcomes like learning and confidence-building. Their case study at Microsoft identified strategies such as the Simple-Complex, Priority-First, and Exploration-Based strategies, which rely heavily on mentoring and peer support to guide new developers. However, mentoring often reduces the productivity of the mentor, placing a significant burden on them. Buchan et al. \cite{buchan2019effectiveteamonboardingagile} observed similar trends in Agile teams, where mentoring activities play a key role in successful onboarding.

The current mentor-focused onboarding process, as described by Britto et al. \cite{britto2017onboarding}, is resource-intensive and less effective in globally distributed projects. Mentoring new developers, especially in remote settings, impacts both mentor and mentee productivity. This paper proposes the Onboarding Buddy as a solution to reduce the mentor burden by providing an autonomous virtual assistant capable of handling routine onboarding queries and supporting the new hires. Onboarding Buddy leverages large language models to offer code explanations, answer project-specific questions, and provide personalized guidance. By automating parts of the onboarding process, this solution aims to increase productivity, reduce stress, and enhance the onboarding experience for both the mentee and the software team the mentee is joining. 
While tools like GitHub Copilot offer code suggestions, they lack the contextual onboarding capabilities that Onboarding Buddy provides, such as tailored, multi-step guidance and integration with project-specific retrieval tools. Unlike ChatGPT plugins, which focus on broad queries, Onboarding Buddy narrows its scope to optimize onboarding efficiency, i.e., we adhere to the code context in the user's question.

\section{Onboarding Buddy}\label{sec:approach}
Our system takes an agent-centric approach to create the Onboarding Buddy. In this section we will elaborate on the LLM orchestration, prompting techniques, and the technologies we used to create the Onboarding Buddy Q\&A capabilities.

In LLM orchestration, a chain is represented by a fixed and rigid sequence of actions. After the chain is implemented, it will act exactly according to the strictly defined steps and without consideration for probable changes in the environment or situation.
For instance, when a chain has been developed to process a user question, it will always execute the very same sequence no matter whether the information is simple or complex, and it cannot adapt if certain data happens to be different from what was anticipated.

Agents behave differently because, in real-time, they would look at the situation with the guidance of an LLM and based on that, decide what and how to perform a certain task. 
An agent executes an action, then observes the result of that action, and dynamically decides what it is going to do next until the desired result is achieved. That can make the agent much more flexible and responsive.  
Through these observations and the LLM-guided RAG, Onboarding Buddy mitigates hallucinations or code mismatch errors by cross-checking final answers with retrieved code snippets before presenting them to the user or to the next agent in the chain.

\begin{figure*}[tb]
    \centering
    \includegraphics[width=0.85\textwidth]{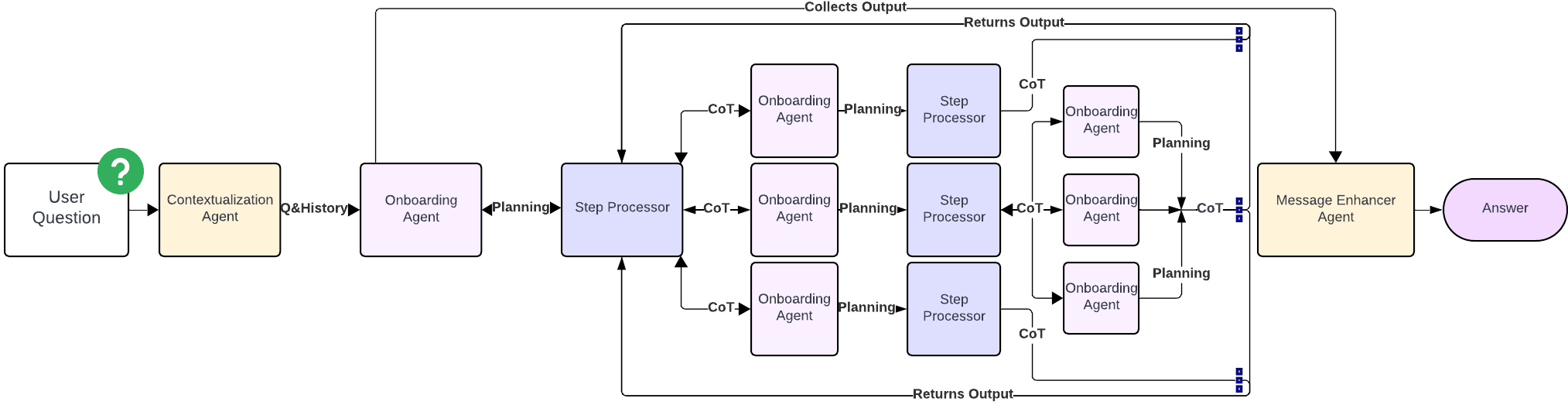}
    \caption{Onboarding Buddy Overall Architecture}
    \label{fig:OnboardingBuddyOverallArchitectureLLM}
\end{figure*}

Our LLM onboarding orchestration implies a chain of agents that collaborate with each other to get a final answer as seen in Figure~\ref{fig:OnboardingBuddyOverallArchitectureLLM}.

\begin{itemize}
    \item \textbf{Contextualization Agent} --- Allows the user interaction to be conversational, taking into account past interactions with the system.
    \item \textbf{Onboarding Agent} --- Allows the user to ask questions about a software project, and provides tailored answers based on the project and user's needs, generating a chain-of-thought step-by-step plan of action.
    \item \textbf{Step Processor} --- A chain-of-thought processor \cite{wei2023chainofthoughtpromptingelicitsreasoning} that orchestrates multiple Onboarding Agents in parallel. It accepts the initial step-by-step plan from an \textbf{Onboarding Agent} and systematically decomposes each high-level step into more granular sub-steps through a recursive approach. For each sub-step, the processor generates partial solutions, and when a sub-task reaches an appropriate level of refinement, it integrates these partial solutions into a final answer.
    \item \textbf{Message Enhancer Agent} --- Allows the system to provide a proper markdown-formatted answer and fix certain inaccuracies in the answer by checking the answer collected from the \textbf{Step Processor} against the codebase.   
\end{itemize}

The \textbf{Onboarding Agent} is the core of our Onboarding Solution, being the agent that creates the planning and the chain-of-thought used to tackle onboarding tasks.

\section{Onboarding Agent}\label{sec:onboardingAgent}
\begin{figure}[bh]
    \centering
    \includegraphics[width=0.4\textwidth]{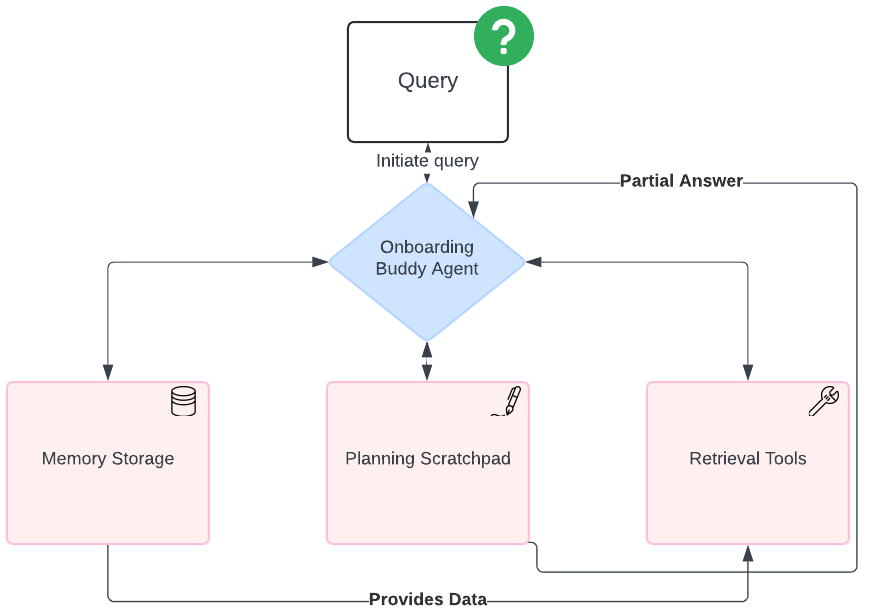}
    \caption{Onboarding Agent Architecture}
    \label{fig:AgentArchitecture}
\end{figure}
This agent has a more complicated structure than the other agents. The 'Onboarding Agent', as seen in Figure~\ref{fig:AgentArchitecture}, is composed of three subcomponents, namely the memory storage, the planning scratchpad, and the retrieval tools.
This agent works iteratively and decides dynamically what actions to take to reach an answer using the planning scratchpad and the retrieval tools. For this, we have implemented a custom chain-of-thought approach (CoT) that will be used in the \texttt{Step Processor}.

The approach we developed was inspired by Google's paper that introduced the concept of chain-of-thought prompting \cite{wei2023chainofthoughtpromptingelicitsreasoning}, where a problem is divided into intermediate reasoning steps, allowing the LLMs to solve more intricate tasks that require a multi-step logic. Instead of getting an immediate answer, the CoT technique forces the model to ``reason'' into human-like steps. On top of that, Amazon Research \cite{zhang2022automaticchainthoughtprompting} elaborated on the same idea from a different angle and developed an automated chain-of-thought generation technique where the manual prompting abilities are minimized. In our approach we used similar ideas, but in an agent-centric approach, where each agent chain in the CoT generation can use external tools such as retrieval tools or code search tools in an exploratory way. For this purpose, each sub-component has its own role in the CoT generation.

\subsection{Memory Storage}

The memory storage comprises 2 sub-components, a database and a blob storage. The database records the user and LLM's past interactions, and the blob storage provides information about the context of the question. In our case, the context is based on details about a software project and its documentation. Furthermore, the blob storage allows the retrieval tools to retrieve relevant information based on the planning scratchpad discussed in subsection \ref{subsec:planningscratchpad}.

\subsection{Retrieval Tools}
\label{sub:retrievalTool}

The retrieval tools allow the agent to access a software project's files in the blob storage and perform semantic searches on its contents. With these tools, we built an advanced, LLM-guided, RAG system. The planning scratchpad presented in subsection \ref{subsec:planningscratchpad}) helps the agent build queries and allows the tools to provide feedback on the quality of the query and search results, enabling the agent to adjust its searches or focus on specific search issues.

The semantic search uses FAISS \cite{faiss}, an in-memory dense-vector database that provides efficient similarity search. Each file, including the documentation, is chunked into documents. Each chunked document is represented by data chunks of 2000 characters with 200 characters overlap. Then each chunk is encoded as a vector using an embedding model (OpenAI \texttt{text-embedding-3-large}~\cite{openai_textembedding3large}).

Each chunked document contains as metadata the GitHub URL that locates the source under the main branch and the path in the project, along with a unique numerical identifier per file.
Based on this data, we make available two tools to the agent that are ``cooperating'' to compose an advanced RAG approach and have the following definitions:
\begin{itemize}
    \item \texttt{retrieve\_missing\_files} - retrieves entire files of source code. The agent can decide to use it only based on the file name information, or if it is guided to do so by the \texttt{retrieve\_relevant\_code\_snippets} tool based on the location available in the metadata.
    \item \texttt{retrieve\_relevant\_code\_snippets} - retrieves chunks of information based on a similarity search using a query to the vector database. This tool has two fallback mechanisms; one handles the cases where the query returns no results, with feedback about the query, and the other one handles the case of repeated queries.
\end{itemize}

At each iteration, the agent can decide to use any of these two tools, but we define the \texttt{retrieve\_missing\_files} as a more computationally expensive tool since it can blindly retrieve files from the storage.
The code snippet retriever returns the top five documents with a similarity score above 0.1, to ensure relevant results. Scores range from 0 (no similarity) to 1 (exact match). The 0.1 threshold filters irrelevant documents but may return zero matches, prompting fallback queries to refine the search.
This, in combination with the agent planning scratchpad presented in Section \ref{subsec:planningscratchpad}, allows the agent to self-adjust the queries and files retrieved, and store this information in the agent planning scratchpad.

\subsection{Planning Scratchpad}\label{subsec:planningscratchpad}

The agent planning scratchpad serves as a dynamic mechanism of internal memory applied by the onboarding agent to track its reasoning process while interacting with different tools, queries, and environments. It works just like a journal or log, where it actually writes down intermediate steps. 

When the agent intends to solve any problem, it may invoke a number of tools presented in the section~\ref{sub:retrievalTool} or various actions sequentially (\textit{e.g.,} retrieve answers from its own knowledge). 
Each subsequent action might produce new data or insights, and these would be stored temporarily in the agent planning scratchpad. For instance, if the agent is trying to find some information in the knowledge base (repository or documentation), gather information from the vector store, or reason out something over various facts, then at each step it writes it in the scratchpad so that it may refer to it later. This allows the agent to recall why it had chosen to perform some particular action or call some specific tool, making its decision process much more transparent and accurate. Additionally, the scratchpad can help potential users with explainability by providing detailed notes on the CoT process of the agent. 

We achieved task chaining using a scratchpad, where each action informs the next to prevent redundancy. For example, an agent can store tool results in the scratchpad for future reference instead of rerunning tools. This also helps guide retrieval tools in the right direction when retrievals fail. If the agent receives new information or needs to change its approach, the scratchpad preserves earlier context while allowing flexibility. Thus, the planning scratchpad serves as a dynamic buffer that considers past decisions and adapts as situations evolve, such as reformulating queries after unsuccessful searches to find the necessary documentation or code snippets.

Moreover, the planning scratchpad is handy during debugging and optimizing agents' workflows. With the contents of the scratchpad, we can identify where an agent might have gone wrong in its reasoning or where an unnecessary call of a tool was made.

In the execution process, the final state of the scratchpad already contains the intermediate results, enabling the agent to assemble a final response that fully answers the user's query while considering the reasoning for each step. 
The answer of this agent returns a step-by-step plan of action with respect to the user's query. For example, if the user asks how to set up a project, the \textbf{Onboarding Agent} provides step-by-step, granular sub-steps of what the user has to perform and how.
\textbf{Onboarding Agent}'s answer is then used further in the chain to the \texttt{Step Processor} that uses multiple instances of the \textbf{Onboarding Agent} in parallel to execute the CoT plan of action created by the initial agent in order to make the steps more precise and granular. Furthermore, the \texttt{Message Enhancer} agent formats, fixes, and beautifies the final answers into a fully readable markdown document.

\section{Study Setup}
To evaluate the usefulness of the Onboarding Buddy, we set up a study. We gathered a sample of eight programmers and assigned each of them three onboarding tasks on an unfamiliar, closed-source project. The first, \textit{``Setup Task''}, requested the user to set up the development environment by cloning the repository, uncommenting specific lines of the configuration file, and verifying that everything was correctly set by registering a mock user to the app and testing specific endpoints. The second task, \textit{``New Payment Option Task''}, required the implementation of a new payment option in the in-application subscription system from which one had to understand an existing payment pipeline and adjust certain code sections. Finally, the \textit{``Questionnaire Duplication Task''} entailed writing a new API endpoint for duplicating questionnaires by implementing duplication logic and writing integration tests to ensure functionality. 

To assess the usefulness of the approach, we collected project telemetry data to track the progress of tasks (\textit{e.g.,} code edits). After completing the tasks, we asked participants to evaluate the helpfulness and ease of onboarding. Both questions required participants to rate the entire system on a scale from 0 to 4, where 0 signifies ``not helpful'' or ``very difficult'', and 4 indicates ``very helpful'' or ``very easy'', respectively. Lastly, we requested open-ended feedback, inviting specific user suggestions for tool enhancements.

Our pilot study involved eight participants on one codebase due to the limitations of the demonstration project. This may affect generalizability. In future evaluations at JetBrains, we aim to include multiple teams with different technology stacks and compare Onboarding Buddy’s performance to off-the-shelf solutions like GitHub Copilot and ChatGPT plugins on similar tasks.

\section{Results}

Seven out of eight participants successfully completed all tasks, with only one participant encountering minor coding errors on a single task. 
Questionnaire results indicated that the perceived helpfulness of the Onboarding Buddy for all tasks had a high mean of \APAstats{3.26}{0.86}. Moreover, the ease-of-onboarding ratings averaged \APAstats{3.0}{0.96}, reflecting a smooth onboarding experience. These mean values represent the combined responses of all study participants across all tasks.
Task completion accuracy was close to 100\% among participants who marked their tasks as finished but also based on the collected telemetry. However, a number of users experienced minor hiccups during their onboarding process, particularly on set-up and questionnaire duplication tasks, which were the hardest tasks we exposed our participants to. 

The most common suggestions were about user experience (UX), including the following: more context awareness by providing more specific code snippets along with line numbers, more user-friendly UI features such as multi-line input and some navigation buttons, and faster response times.
Although not a formal benchmark, participants who used mainstream AI coding tools in their daily workflow noted in their feedback that Onboarding Buddy gave more context-specific advice, particularly on setup tasks. This suggests potential advantages over generic code assistants, though rigorous A/B testing shall be done.

\section{Conclusion}
Overall, the Onboarding Buddy is an easy-to-use and useful tool that helps users in performing development tasks. 
This design is modular and adaptive, capable of efficiently solving problems using techniques like chain-of-thought reasoning and retrieval tools. However, addressing the identified shortcomings, enhancing technical reliability, and optimizing the user interface will significantly improve the overall usability and utility of the Onboarding Buddy. Refining these aspects will make the Onboarding Buddy capable of providing a smoother and more effective onboarding experience, thereby further enhancing user satisfaction and productivity.
In future work, we will incorporate more robust validation checks (e.g., hallucination mitigation), and benchmark Onboarding Buddy against GitHub Copilot and ChatGPT-based solutions in varied (enterprise) settings.
Regarding scalability, future versions will involve live project updates, reducing reliance on pre-encoded project data.

\section{Acknowledgment}
This work was conducted as part of the AI for Software Engineering (AI4SE) collaboration between JetBrains and Delft University of Technology. The authors gratefully acknowledge the financial support provided by JetBrains, which made this research possible.

\bibliographystyle{ACM-Reference-Format}
\bibliography{main}


\begin{thebibliography}{16}


\ifx \showCODEN    \undefined \def \showCODEN     #1{\unskip}     \fi
\ifx \showISBNx    \undefined \def \showISBNx     #1{\unskip}     \fi
\ifx \showISBNxiii \undefined \def \showISBNxiii  #1{\unskip}     \fi
\ifx \showISSN     \undefined \def \showISSN      #1{\unskip}     \fi
\ifx \showLCCN     \undefined \def \showLCCN      #1{\unskip}     \fi
\ifx \shownote     \undefined \def \shownote      #1{#1}          \fi
\ifx \showarticletitle \undefined \def \showarticletitle #1{#1}   \fi
\ifx \showURL      \undefined \def \showURL       {\relax}        \fi
\providecommand\bibfield[2]{#2}
\providecommand\bibinfo[2]{#2}
\providecommand\natexlab[1]{#1}
\providecommand\showeprint[2][]{arXiv:#2}

\bibitem[Al-Kaswan et~al\mbox{.}(2023)]%
        {al2023extending}
\bibfield{author}{\bibinfo{person}{Ali Al-Kaswan}, \bibinfo{person}{Toufique Ahmed}, \bibinfo{person}{Maliheh Izadi}, \bibinfo{person}{Anand~Ashok Sawant}, \bibinfo{person}{Premkumar Devanbu}, {and} \bibinfo{person}{Arie van Deursen}.} \bibinfo{year}{2023}\natexlab{}.
\newblock \showarticletitle{Extending source code pre-trained language models to summarise decompiled binaries}. In \bibinfo{booktitle}{\emph{2023 IEEE International Conference on Software Analysis, Evolution and Reengineering (SANER)}}. IEEE, \bibinfo{pages}{260--271}.
\newblock


\bibitem[Britto et~al\mbox{.}(2017)]%
        {britto2017onboarding}
\bibfield{author}{\bibinfo{person}{Ricardo Britto}, \bibinfo{person}{Daniela~S. Cruzes}, \bibinfo{person}{Darja Smite}, {and} \bibinfo{person}{Aivars Sablis}.} \bibinfo{year}{2017}\natexlab{}.
\newblock \showarticletitle{Onboarding software developers and teams in three globally distributed legacy projects: A multicase study}.
\newblock \bibinfo{journal}{\emph{Journal of Software: Evolution and Process}} \bibinfo{volume}{30}, \bibinfo{number}{4} (\bibinfo{date}{11} \bibinfo{year}{2017}).
\newblock


\bibitem[Buchan et~al\mbox{.}(2019)]%
        {buchan2019effectiveteamonboardingagile}
\bibfield{author}{\bibinfo{person}{Jim Buchan}, \bibinfo{person}{Stephen MacDonell}, {and} \bibinfo{person}{Jennifer Yang}.} \bibinfo{year}{2019}\natexlab{}.
\newblock \bibinfo{title}{Effective team onboarding in Agile software development: techniques and goals}.
\newblock
\showeprint[arxiv]{1907.10206}~[cs.SE]
\urldef\tempurl%
\url{https://arxiv.org/abs/1907.10206}
\showURL{%
\tempurl}


\bibitem[Deljouyi et~al\mbox{.}(2024)]%
        {deljouyi2024leveraging}
\bibfield{author}{\bibinfo{person}{Amirhossein Deljouyi}, \bibinfo{person}{Roham Koohestani}, \bibinfo{person}{Maliheh Izadi}, {and} \bibinfo{person}{Andy Zaidman}.} \bibinfo{year}{2024}\natexlab{}.
\newblock \showarticletitle{Leveraging large language models for enhancing the understandability of generated unit tests}.
\newblock \bibinfo{journal}{\emph{arXiv preprint arXiv:2408.11710}} (\bibinfo{year}{2024}).
\newblock


\bibitem[Ernst and Robillard(2023)]%
        {ernst2023documentation}
\bibfield{author}{\bibinfo{person}{Neil~A. Ernst} {and} \bibinfo{person}{Martin~P. Robillard}.} \bibinfo{year}{2023}\natexlab{}.
\newblock \showarticletitle{A study of documentation for software architecture}.
\newblock \bibinfo{journal}{\emph{Empirical Software Engineering}} \bibinfo{volume}{28}, \bibinfo{number}{122} (\bibinfo{date}{September} \bibinfo{year}{2023}), \bibinfo{pages}{1--32}.
\newblock
\href{https://doi.org/10.1007/s10664-023-10347-2}{doi:\nolinkurl{10.1007/s10664-023-10347-2}}
\newblock
\shownote{Accepted 25 May 2023, Published 15 September 2023}.


\bibitem[Ionescu et~al\mbox{.}(2025a)]%
        {onboardingbuddy2025video}
\bibfield{author}{\bibinfo{person}{Andrei-Cristian Ionescu}, \bibinfo{person}{Sergey Titove}, {and} \bibinfo{person}{Maliheh Izadi}.} \bibinfo{year}{2025}\natexlab{a}.
\newblock \bibinfo{title}{Onboarding Buddy - Demonstration Video}.
\newblock \bibinfo{howpublished}{\url{https://youtu.be/V18mjpziCvo}}.
\newblock
\newblock
\shownote{Accessed: 2025-03-27}.


\bibitem[Ionescu et~al\mbox{.}(2025b)]%
        {onboardingbuddy2025software}
\bibfield{author}{\bibinfo{person}{Andrei-Cristian Ionescu}, \bibinfo{person}{Sergey Titove}, {and} \bibinfo{person}{Maliheh Izadi}.} \bibinfo{year}{2025}\natexlab{b}.
\newblock \bibinfo{title}{Onboarding Buddy - Tool and Source Code}.
\newblock \bibinfo{howpublished}{\url{https://onboarding.software/register?colab=true}}.
\newblock
\newblock
\shownote{Accessed: 2025-03-27}.


\bibitem[Izadi et~al\mbox{.}(2024)]%
        {izadi2024language}
\bibfield{author}{\bibinfo{person}{Maliheh Izadi}, \bibinfo{person}{Jonathan Katzy}, \bibinfo{person}{Tim Van~Dam}, \bibinfo{person}{Marc Otten}, \bibinfo{person}{Razvan~Mihai Popescu}, {and} \bibinfo{person}{Arie Van~Deursen}.} \bibinfo{year}{2024}\natexlab{}.
\newblock \showarticletitle{Language models for code completion: A practical evaluation}. In \bibinfo{booktitle}{\emph{Proceedings of the IEEE/ACM 46th International Conference on Software Engineering}}. \bibinfo{pages}{1--13}.
\newblock


\bibitem[Johnson et~al\mbox{.}(2017)]%
        {faiss}
\bibfield{author}{\bibinfo{person}{Jeff Johnson}, \bibinfo{person}{Matthijs Douze}, {and} \bibinfo{person}{Herv{\'e} J{\'e}gou}.} \bibinfo{year}{2017}\natexlab{}.
\newblock \bibinfo{title}{{Billion-scale similarity search with GPUs}}.
\newblock \bibinfo{howpublished}{\url{https://github.com/facebookresearch/faiss}}.
\newblock
\newblock
\shownote{Accessed: 2024-09-27}.


\bibitem[Ju et~al\mbox{.}(2021)]%
        {ju2021case}
\bibfield{author}{\bibinfo{person}{An Ju}, \bibinfo{person}{Hitesh Sajnani}, \bibinfo{person}{Scot Kelly}, {and} \bibinfo{person}{Kim Herzig}.} \bibinfo{year}{2021}\natexlab{}.
\newblock \bibinfo{title}{A Case Study of Onboarding in Software Teams: Tasks and Strategies}.
\newblock
\showeprint[arxiv]{2103.05055}~[cs.SE]


\bibitem[Lewis et~al\mbox{.}(2021)]%
        {lewis2021retrievalaugmentedgenerationknowledgeintensivenlp}
\bibfield{author}{\bibinfo{person}{Patrick Lewis}, \bibinfo{person}{Ethan Perez}, \bibinfo{person}{Aleksandra Piktus}, \bibinfo{person}{Fabio Petroni}, \bibinfo{person}{Vladimir Karpukhin}, \bibinfo{person}{Naman Goyal}, \bibinfo{person}{Heinrich Küttler}, \bibinfo{person}{Mike Lewis}, \bibinfo{person}{Wen tau Yih}, \bibinfo{person}{Tim Rocktäschel}, \bibinfo{person}{Sebastian Riedel}, {and} \bibinfo{person}{Douwe Kiela}.} \bibinfo{year}{2021}\natexlab{}.
\newblock \bibinfo{title}{Retrieval-Augmented Generation for Knowledge-Intensive NLP Tasks}.
\newblock
\showeprint[arxiv]{2005.11401}~[cs.CL]
\urldef\tempurl%
\url{https://arxiv.org/abs/2005.11401}
\showURL{%
\tempurl}


\bibitem[MacNeil et~al\mbox{.}(2023)]%
        {MacNeil2023Experiences}
\bibfield{author}{\bibinfo{person}{Stephen MacNeil}, \bibinfo{person}{Andrew Tran}, \bibinfo{person}{Arto Hellas}, \bibinfo{person}{Joanne Kim}, \bibinfo{person}{Sami Sarsa}, \bibinfo{person}{Paul Denny}, \bibinfo{person}{Seth Bernstein}, {and} \bibinfo{person}{Juho Leinonen}.} \bibinfo{year}{2023}\natexlab{}.
\newblock \showarticletitle{Experiences from Using Code Explanations Generated by Large Language Models in a Web Software Development E-Book}. In \bibinfo{booktitle}{\emph{Proceedings of the 54th ACM Technical Symposium on Computer Science Education V. 1}}. ACM.
\newblock


\bibitem[OpenAI(2023)]%
        {openai_textembedding3large}
\bibfield{author}{\bibinfo{person}{OpenAI}.} \bibinfo{year}{2023}\natexlab{}.
\newblock \bibinfo{title}{text-embedding-3-large}.
\newblock \bibinfo{howpublished}{\url{https://platform.openai.com/docs/guides/embeddings}}.
\newblock
\newblock
\shownote{Language model}.


\bibitem[Sergeyuk et~al\mbox{.}(2024)]%
        {sergeyuk2024design}
\bibfield{author}{\bibinfo{person}{Agnia Sergeyuk}, \bibinfo{person}{Ekaterina Koshchenko}, \bibinfo{person}{Ilya Zakharov}, \bibinfo{person}{Timofey Bryksin}, {and} \bibinfo{person}{Maliheh Izadi}.} \bibinfo{year}{2024}\natexlab{}.
\newblock \showarticletitle{The Design Space of in-IDE Human-AI Experience}.
\newblock \bibinfo{journal}{\emph{arXiv preprint arXiv:2410.08676}} (\bibinfo{year}{2024}).
\newblock


\bibitem[Wei et~al\mbox{.}(2023)]%
        {wei2023chainofthoughtpromptingelicitsreasoning}
\bibfield{author}{\bibinfo{person}{Jason Wei}, \bibinfo{person}{Xuezhi Wang}, \bibinfo{person}{Dale Schuurmans}, \bibinfo{person}{Maarten Bosma}, \bibinfo{person}{Brian Ichter}, \bibinfo{person}{Fei Xia}, \bibinfo{person}{Ed Chi}, \bibinfo{person}{Quoc Le}, {and} \bibinfo{person}{Denny Zhou}.} \bibinfo{year}{2023}\natexlab{}.
\newblock \bibinfo{title}{Chain-of-Thought Prompting Elicits Reasoning in Large Language Models}.
\newblock
\showeprint[arxiv]{2201.11903}~[cs.CL]
\urldef\tempurl%
\url{https://arxiv.org/abs/2201.11903}
\showURL{%
\tempurl}


\bibitem[Zhang et~al\mbox{.}(2022)]%
        {zhang2022automaticchainthoughtprompting}
\bibfield{author}{\bibinfo{person}{Zhuosheng Zhang}, \bibinfo{person}{Aston Zhang}, \bibinfo{person}{Mu Li}, {and} \bibinfo{person}{Alex Smola}.} \bibinfo{year}{2022}\natexlab{}.
\newblock \bibinfo{title}{Automatic Chain of Thought Prompting in Large Language Models}.
\newblock
\showeprint[arxiv]{2210.03493}~[cs.CL]
\urldef\tempurl%
\url{https://arxiv.org/abs/2210.03493}
\showURL{%
\tempurl}


\end{thebibliography}

\end{document}